\documentclass{article}
\usepackage[left=1in,right=1in,bottom=1in,top=1in]{geometry}
\usepackage{amsmath}
\usepackage{amssymb}
\usepackage{amsthm}
\usepackage{hyperref}

\title{Errata for \\
``Global existence and scattering for the nonlinear Schr\"odinger equation on Schwarzschild manifolds'',\\
 ``Semilinear wave equations on the Schwarzschild manifold I: Local Decay Estimates'', and \\
``The wave equation on the Schwarzschild metric II: Local Decay for the spin 2 Regge Wheeler equation''.}
\author{P. Blue, A. Soffer}
\date{6 September 2006}

\newcommand{\gasblong}{g_{\alpha_l,\sigma,b}}
\newcommand{\gammaasblong}{\gamma_{\alpha_l,\sigma,b}}
\newcommand{\gasb}{g}
\newcommand{\gammaasb}{\gamma}

\newcommand{\InverseCubicClean}{\frac{1}{(1+\rs^2)^{\sigma+2}}}

\newcommand{\InverseLin}{\frac{b}{(1+b|\rsas|)^\sigma}}
\newcommand{\InverseSquare}{\frac{b^2\sigma\sgn(\rsas)}{(1+b|\rsas|)^{\sigma+1}}}
\newcommand{\InverseCubic}{\frac{b^3\sigma(\sigma+1)}{(1+b|\rsas|)^{\sigma+2}}}

\newcommand{\InverseLinWO}{\frac{1}{(1+b|\rsas|)^\sigma}}
\newcommand{\InverseCubicWO}{\frac{1}{(1+b|\rsas|)^{\sigma+2}}}

\newcommand{\rs}{{r_*}}
\newcommand{\as}{{(\alpha_l)_*}}
\newcommand{\rsas}{{\rs-\as}}
\newcommand{\dr}{\partial_\rs}
\newcommand{\drr}{\partial_\rs^2}
\newcommand{\dt}{\partial_t}

\newcommand{\starman}{\mathcal{M}}
\newcommand{\dThree}{d\rs d^2\omega}

\newcommand{\sgn}{\text{sgn}}
\newcommand{\Reals}{\mathbb{R}}

\newcommand{\sfn}{u}

\newcommand{\HS}{H_{\starman}}
\newcommand{\HRW}{H_{\text{RW}}}
\newcommand{\Htwop}{H_{2'}}

\newcommand{\potl}{V_{\text{eff}}}

\begin{document}
\maketitle

\begin{abstract}
In \cite{LabaSoffer}, \cite{BlueSoffer1}, \cite{BlueSoffer2}, local decay estimates were proven for the (decoupled) Schr\"odinger, wave, and Regge-Wheeler equations on the Schwarzschild manifold, using commutator methods. Here, we correct a step in the commutator argument. The corrected argument works either for radial semilinear equations or general linear equations. This recovers the results in \cite{LabaSoffer} and \cite{BlueSoffer2}, but does not recover the non radial, large data, semilinear result asserted in \cite{BlueSoffer1}. 
\end{abstract}

In \cite{LabaSoffer}, \cite{BlueSoffer1}, \cite{BlueSoffer2}, various equations are considered on the Schwarzschild manifold, with the aim of making progress towards understanding the stability of the Schwarzschild solution. These papers use vector field methods and similar commutator arguments to get local decay results. We report here that these papers all contain an error in the calculation of a commutator of the form $i[-\drr, (1/2)(g(-i\dr)+(-i\dr) g)]$ involving the multiplier $(1/2)(g(-i\dr)+(-i\dr) g)$. These papers have been followed by further work by ourselves and others \cite{BlueSoffer3}, \cite{BSterbenz}, \cite{DafermosRodnianski}, using additional vector field arguments to prove stronger decay results for wave equations. Local decay arguments are a necessary part of these results, and both \cite {BSterbenz} and \cite{DafermosRodnianski} provide corrected arguments at the analogous stage. Interesting decay results have also been proven using very different methods in \cite{FinsterKamranSmollerYau}. Here, we have modified the multiplier, $(1/2)(g(-i\dr)+(-i\dr)g)$. \footnote{We use the modification first suggested by J. Sterbenz.}

By local decay estimates, we mean space time integral estimates of the form, for $\sigma>1$,  
\begin{align}
\label{eLocalDecay}
\int_1^\infty \int_{\Reals\times S^2} \InverseCubicClean |\sfn|^2 d\rs d^2\omega dt \leq& C( E + \|\sfn_0\|^2) ,
\end{align}
where $u$ is the solution to the relevant equation, $(t,\rs,\omega)\in\Reals\times\starman=\Reals\times\Reals\times S^2$ is the Regge-Wheeler co-ordinate system for the exterior region of the Schwarzschild solution, $E$ is the conserved energy of the solution, and $\|u_0\|$ is the $L^2$ norm of the initial data. In fact, we can also control weighted integrals of derivatives as well, which we take advantage of in \cite{BlueSoffer3}. The equations considered in \cite{LabaSoffer}, \cite{BlueSoffer1}, and \cite{BlueSoffer2} are, respectively, 
\begin{align}
i\dt \sfn =& \HS \sfn + \lambda r^{1-p}|\sfn|^{p-1} \sfn , \label{eSchro}\\
-\ddot{\sfn} =& \HS \sfn + \lambda r^{1-p} |\sfn|^{p-1} \sfn, \label{eWave}\\
-\ddot{\sfn} =& \HRW \sfn . \label{eRW}
\end{align}
These are, respectively, the (non-linear) Schr\"odinger equation, the wave equation, and the Regge-Wheeler equation. The initial data for \eqref{eSchro} is $\sfn(1)=\sfn_0$, and the initial date for \eqref{eWave} and \eqref{eRW} is $\sfn(1)=\sfn_0$, $\dot{\sfn}(1)=\sfn_1$. For \eqref{eSchro} in \cite{LabaSoffer}, the defocusing case $\lambda\geq0$ with radial initial data is considered; the range of $p$ values is explained in \cite{LabaSoffer}, but is included in $p\geq3$ and requires $p>4$ for further results. We can recover the estimate \eqref{eLocalDecay} for $p$ from $3$ to a little more than $4$. For \eqref{eWave} in \cite{BlueSoffer1}, we are not able to recover the previous claimed results. For either the cubic, defocusing case $p=3$, $\lambda\geq0$ with radial initial data or the linear case $\lambda=0$ with general (non radial) initial data, we can recover the estimate \eqref{eLocalDecay}, but we can not for the non-linear equation with general data. General data is considered for the Regge-Wheeler equation. For the Schr\"odinger equation, $u$ is a complex valued function, and for the wave and Regge-Wheeler equations, it is real valued. Because of the structure of the Regge-Wheeler equation, the minimum spherical harmonic parameter is $l=2$.

The Hamiltonian terms are
\begin{align*}
\HS=& H_1 + H_2 + H_3 , \\
\HRW=& H_1 + \Htwop + H_3 , \\
H_1=& -\drr , \\
H_2=& V = \frac{2M}{r^3} \left( 1- \frac{2M}{r}\right) , \\
\Htwop=& -3 V , \\
H_3 =& V_L (-\Delta_S^2) , \\
V_L=& \frac{1}{r^2}\left( 1 - \frac{2M}{r}\right) . 
\end{align*} 
The quantity $r$ is an alternate radial co-ordinate which we treat as a function of $\rs$ such that $dr/d\rs = (1-2M/r)$. 

Following the argument in \cite{LabaSoffer}, \cite{BlueSoffer1}, and \cite{BlueSoffer2}, to prove the local decay estimate \eqref{eLocalDecay}, in the linear case, it is sufficient to find an operator $\gamma$ such that for any sufficiently smooth $\sfn$, $\|\gamma \sfn\| \leq C_1 \| \dr u \| + C_2 \| (1+\rs^2)^{-\sigma/2} u \|$, and
\begin{align}
\label{eLinCommutator}
\langle\sfn,i[H,\gamma]\sfn\rangle \geq& C \int_\starman \InverseCubicClean |\sfn|^2 d\rs d^2\omega . 
\end{align}
For the non-linear case, the following additional estimate is sufficient
\begin{align} 
\label{eNLCommutator}
\langle\sfn, i[\lambda r^{1-p}|u|^{p-1},\gamma]\sfn\rangle \geq C \int_\starman \left(1-\frac{2M}{r}\right) r^{-p}|\sfn|^{p+1} d\rs d^2\omega .
\end{align}

The estimate \eqref{eLinCommutator} can be proven on each spherical harmonic using $\gamma$ of the form $(-i/2)(g\dr + \dr g)$, with $g$ centred at the peak of the effective potential. To extend the result to general data, it is sufficient to prove that the estimate holds uniformly in the spherical harmonic parameter. Finally, we use an integration by parts argument to prove \eqref{eNLCommutator} when only one value of $g$ has been used. Because the effective potentials have different maxima, we have had to use different $g$'s on each spherical harmonic and have not been able to recover the result for non-linear equations with general initial data in \cite{BlueSoffer1}. 

For the Schr\"odinger and wave equations, the effective potential on each spherical harmonic is 
\begin{align*}
\potl:= V + l(l+1) V_L .
\end{align*}
The derivative is 
\begin{align*}
\potl' = -\frac{2}{r^5}\left( l(l+1)r^2 - (l(l+1)-1) 3M r -8M^2 \right) \left( 1- \frac{2M}{r}\right).
\end{align*}
In the range $r>2M$, this has a unique maximum at the following value of $r$, 
\begin{align*}
\alpha_l=& \frac{(3l(l+1)-3)M + \sqrt{ (3-3l(l+1))^2M^2+32 l(l+1)M^2}}{2l(l+1)} .
\end{align*}
For $l=0$ the unique root is at $r=8M/3$. For $l>0$, there is a second, negative root which can be ignored. Let $\as$ be the value of $\rs$ corresponding to $r=\alpha_l$. As $l\rightarrow\infty$, $\alpha_l\rightarrow3M$ and $\potl/(l(l+1))\rightarrow V_L$, so that the second derivative of $\potl$ at $\as$ is uniformly bounded below. 

For the Regge-Wheeler equation, the effective potential, its derivative, and the value of $r$ corresponding to the unique maximum are 
\begin{align*}
\potl =& -3V + l(l+1) V_L , \\
\potl' =& -\frac{2}{r^5}\left( l(l+1)r^2 - (l(l+1)+3) 3M r + 24M^2 \right) \left( 1- \frac{2M}{r}\right), \\
\alpha_l=& \frac{(3l(l+1)+9)M + \sqrt{ (9+3l(l+1))^2M^2-96l(l+1)M^2}}{2l(l+1)} .
\end{align*}
Again, there is a second root, but for $l=2$, this root is less than $2M$, and as $l$ increases, the second root decreases, so they can be ignored. Again, the maxima converge, and the second derivatives at the critical point are uniformly bounded from below, so the estimate on each spherical harmonic can be summed, to provide a uniform estimate. 

Given $H=-\drr + \potl$, where $\potl$ has a unique maximum at $\as$, for parameters $\sigma>1$ and $b>0$, let
\begin{align*}
\gasblong=& \int_0^{b(\rs-\as)} \frac{1}{(1+|x|)^\sigma} dx ,\\
\gammaasblong=& -i(\gasblong \dr + \frac12\gasblong') . 
\end{align*}
We will typically denote these $\gasb$ and $\gammaasb$ for simplicity. Since $\sigma>1$, $\gasblong$ is bounded, and $\| \gammaasblong \sfn\| \leq C_1 \|\dr\sfn\|+ C_2 \| (1+\rs^2)^{-\sigma/2} u \|$ uniformly in $\alpha_l$ for each $\sigma$ and $b$. 

From the definition, 
\begin{align*}
\gasb'  =& \InverseLin , \\
\gasb'' =& -\InverseSquare , \\
\gasb'''=& -2b^2\sigma\delta_\as + \InverseCubic .
\end{align*}

The commutator is
\begin{align*}
\langle \sfn,i[-\drr + \potl,\gammaasb]\sfn\rangle 
=& \int_\starman 2\gasb' (\dr \sfn)^2 -\frac12 \gasb'''\sfn^2 - \gasb \potl' |\sfn|^2 \dThree .
\end{align*}
Since $\potl$ goes from increasing to decreasing at $\as$, the last term is strictly positive. 

Let $\chi$ be a smooth, compactly supported function, which is identically one in a neighbourhood of $\as$. This can be used to estimate $\sfn(\as)$ on each spherical harmonic. We use a normalised measure on the sphere, so that $\int_{S^2}d^2\omega =1$, and use $|u(\as)|^2$ as abbreviated notation for $\int_{S^2} |u(\as,\omega)|^2 d^2\omega$. 
\begin{align*}
0
=& - \int_\starman \dr ((\rsas) \chi |\sfn|^2) \dThree \\
\int_\starman \chi |\sfn|^2 \dThree
=& -\int_\starman (\rsas)\chi' |\sfn|^2 - \int_\starman (\rsas)\chi 2\Re(\bar{\sfn}\dr \sfn) \dThree \\
\leq& \int_\starman (|(\rsas)\chi'| + (\rsas)^2 \chi)|\sfn|^2 \dThree 
+ \int_\starman \chi |\dr \sfn|^2 \dThree .
\end{align*}
\begin{align*}
|\sfn(\as)|^2
=& -\int_{S^2}\int_0^\infty \dr (\chi |\sfn|^2) \dThree \\
=& -\int_{S^2}\int_0^\infty \chi' |\sfn|^2 \dThree - \int_{S^2} \int_0^\infty \chi2\Re(\bar{\sfn}\dr \sfn)\dThree \\
\leq& \int_{S^2} \int_0^\infty (|\chi'|+\chi) |\sfn|^2 \dThree 
+ \int_{S^2} \int_0^\infty \chi |\dr \sfn|^2 \dThree
\end{align*}
Applying the same argument for $\rs<\as$ and summing, 
\begin{align}
2|\sfn(\as)|^2
\leq& \int_\starman (|\chi'| + |(\rsas)\chi'| + (\rsas)^2 \chi)|\sfn|^2 \dThree 
+ \int_\starman 2 \chi |\dr \sfn|^2 \dThree . 
\label{eDelta}
\end{align}

A similar integration by parts argument can be applied to control $(1+b|\rsas|)^{-\sigma-2}$ by other terms appearing in the commutator. 
\begin{align}
0
=& - \int_\starman \dr (\gasb'' |\sfn^2|) \dThree \nonumber\\
\int_\starman \InverseCubic |\sfn|^2 \dThree
=& 2b^2\sigma |\sfn(\as)|^2 - \int_\starman \InverseSquare 2\Re(\bar{\sfn}\dr \sfn) \dThree \nonumber\\
\leq& 2b^2\sigma |\sfn(\as)|^2 + \frac12 \int_\starman \InverseCubic |\sfn|^2 \dThree \nonumber\\
&+\frac12\frac{4\sigma}{\sigma+1}\int_\starman \InverseLin |\dr \sfn|^2 \dThree \nonumber\\
\int_\starman \InverseCubic |\sfn|^2 \dThree
 =& 4b^2\sigma |\sfn(\as)|^2
+ \frac{4\sigma}{\sigma+1} \int_\starman \InverseLin |\dr \sfn|^2 \dThree .
\label{eInvCub} 
\end{align}

Define the following abbreviated notation, 
\begin{align*}
A=& \int_\starman \InverseLinWO |\dr \sfn|^2 \dThree , \\
B=& \int_\starman \InverseCubicWO |\sfn|^2 \dThree , \\
C=& |\sfn(\as)|^2 , \\
D=& \int_\starman -\gasb \potl' |\sfn|^2 \dThree , \\
E=& \int_\starman 2\chi|\dr \sfn|^2 \dThree , \\
F=& \int_\starman (|\chi'| + |(\rsas)\chi'| + (\rsas)^2 \chi)|\sfn|^2 \dThree . 
\end{align*}
In abbreviated notation, 
\begin{align}
\label{eDeltaAb}
2C\leq& E + F , \\
\label{eInvCubAb}
b^3\sigma(\sigma+1) B \leq& 4b^2\sigma C + \frac{4\sigma}{\sigma+1} b A .
\end{align}

The commutator becomes
\begin{align*}
\langle \sfn,[-\drr+\potl,\gammaasb]\sfn\rangle 
=& 2bA -\frac{b^3\sigma(\sigma+1)}{2} B + b^2\sigma C + D \\
\geq& 2bA - \frac{2\sigma}{\sigma+1}bA -b^2\sigma C +D \\
\geq& \frac{2}{\sigma+1}bA -\frac{b^2\sigma}{2} E -\frac{b^2\sigma}{2}F + D
\end{align*}

The parameter $b$ can be taken sufficiently sufficiently small so that for some constant $c$, 
\begin{align}
\label{ebsmall}
\frac{2}{\sigma+1}bA -\frac{b^2\sigma}{2} E \geq& c A , \\
-\frac{b^2\sigma}{2}F + D                   \geq& c F .
\label{ebsmall2}
\end{align} 
Thus the commutator controls factors of $A$ and $F$. By \eqref{eDeltaAb}, this controls the $\delta$ function $C$, and by \eqref{eInvCubAb}, this controls the local decay term, $B$. Therefore, we get there's a constant so that \eqref{eLinCommutator} holds on each spherical harmonic. 

On each spherical harmonic, we have proven
\begin{align*}
\langle \sfn,[H,\gammaasblong]\sfn\rangle \geq& c' b^3\sigma(\sigma+1)\int \InverseCubicWO |\sfn|^2 d\rs d^2\omega ,
\end{align*}
where $c'$ and $b$ depend on the second derivative of $\potl$ at $\as$ through equation \eqref{ebsmall2}. Since the $\alpha_l$ converge to $3M$, and $V_L'$ vanishes linearly at $r=3M$, for sufficiently large $l$, at $\as$, $l(l+1)V_L''$ dominates $V''$, and $\potl''$ is uniformly bounded below. For each of the finite number of remaining $l$, $\potl'$ vanishes linearly at $\as$. Since the second derivative of the effective potential is uniformly bounded below for $\HS$ and $\HRW$ at $\as$ on each spherical harmonic, there is a uniform choice of $b$. Thus, we have a uniform constant $C$ such that
\begin{align*} 
\langle \sfn,[H,\gammaasblong]\sfn\rangle \geq& 
C \int (1+|r_*|)^{-\sigma-2} |\sfn|^2 d\rs d^2\omega
\end{align*}
This can be summed across the spherical harmonics to prove \eqref{eLinCommutator}. 

Equation \eqref{eNLCommutator} can be proven by an integration by parts argument
\begin{align*}
\langle \sfn, i[\lambda r^{1-p} |\sfn|^{p-1},\gammaasblong]\sfn \rangle
=& -\int_\starman |\sfn|^2 g \dr( \lambda r^{1-p} |\sfn|^{p-1} ) d\rs d^2\omega \\
=& -\lambda\int_\starman g r^2 (1-\frac{2M}{r})^{-\frac{2}{p-1}} \frac{p-1}{p+1} \dr (r^{-1-p} |\sfn|^{p+1} (1-\frac{2M}{r})^{\frac{p+1}{p-1}}) d\rs d^2\omega \\
=& \lambda \frac{p-1}{p+1} \int_\starman \dr(g r^2 (1-\frac{2M}{r})^{-\frac{2}{p-1}})r^{-1-p} |\sfn|^{p+1} (1-\frac{2M}{r})^{\frac{p+1}{p-1}} d\rs d^2\omega 
\end{align*}
The derivative in the integral can be reduced to 
\begin{align*}
\dr(g r^2 (1-\frac{2M}{r})^{-\frac{2}{p-1}})
=& (1-\frac{2M}{r})^{-\frac{2}{p-1}} (2rg(1-\frac{2M}{r}(1+\frac{1}{p-1})) + r^2g')
\end{align*}
The coefficient of $|\sfn|^{p+1}$ decays like $(1-2M/r)$ as $\rs\rightarrow-\infty$ and like $r^{-p}$ as $\rs\rightarrow\infty$. Since the coefficient is continuous, it is sufficient to show that it is strictly positive. The quantity $r^2 g'$ is positive. For $H_2+H_3$ with radial data, the value of $r$ corresponding to $\rs=\as$ is $r=8M/3$. For $4\geq p\geq3$, $ g(1-(2M/r)(1+1/(p-1))$ is negative for $r$ in $(8M/3, 2M(1+1/(p-1))\subset(8M/3,3M)$ (The condition $4\geq p$ guarantees $2M(1+1/(p-1))\geq 8M/3$). For $b$ sufficiently large, in this region, $g'\sim1$, and $|g| \leq 4(r-8M/3)$. The quantity that needs to be controlled is bounded below by 
\begin{align*}
8(r-\frac{8M}{3})(r-3M) +r^2 = \frac{1}{3}(27r^2-8(17)Mr + 3(64) M^2)
\end{align*}
The discriminant for the roots of this is $8^2(17)^2-4(27)3(64)=8^2(17^2-18^2)<0$, so this has no roots and is always positive. By continuity, we can extend this to values of $p$ slightly greater than $4$ to recover the scattering results for some values of $p$ in \cite{LabaSoffer}. 

\vspace{.5in}
\noindent{\bf Acknowledgement}

We would like to thank I. Rodinaski for pointing out for the first time the gap in the local decay computation.

\end{document}